\pdfoutput=1
\documentclass[aps,prd,amsmath,floats,floatfix, twocolumn,
superscriptaddress,nofootinbib,showpacs,longbibliography]{revtex4-1}
\usepackage[T1]{fontenc}
\usepackage[utf8]{inputenc}
\usepackage{txfonts}
\usepackage{verbatim}
\usepackage[dvipsnames, usenames]{xcolor}
\definecolor{linkcolor}{rgb}{0.0,0.3,0.5}
\usepackage[hypertexnames=false, unicode, colorlinks=true, linkcolor=linkcolor,
citecolor=linkcolor, filecolor=linkcolor,urlcolor=linkcolor,
pdfusetitle]{hyperref}
\usepackage[all]{hypcap}
\usepackage{graphicx}
\usepackage{xspace}
\usepackage{amssymb}
\usepackage{microtype}
\usepackage{subfigure}

\usepackage{url}

\usepackage[english]{babel}
\usepackage{blindtext}

\usepackage[normalem]{ulem} %for \sout
\usepackage{bm} % boldmath
\usepackage{mathrsfs}
\usepackage{xspace} % for \xspace

\DeclareMathAlphabet{\mathpzc}{OT1}{pzc}{m}{it}
\usepackage[nomessages]{fp}

\newcommand{\bl}{\ensuremath{\boldsymbol{\lambda}}\xspace}
\newcommand{\sk}[1]{}

\begin{document}
\title{Universal phenomenological relations between spherical harmonic modes\\ in non-precessing eccentric binary black hole merger waveforms}
\newcommand{\KITP}{\affiliation{Kavli Institute for Theoretical Physics, University of California Santa Barbara, Kohn Hall, Lagoon Rd, Santa Barbara, CA 93106}} 
\newcommand{\TAPIR}{\affiliation{Theoretical AstroPhysics Including Relativity and Cosmology, California Institute of Technology, Pasadena, California, USA}}

\author{Tousif Islam}
\email{tousifislam@kitp.ucsb.edu}
\KITP
\TAPIR

\author{Tejaswi Venumadhav}
\affiliation{\mbox{Department of Physics, University of California at Santa Barbara, Santa Barbara, CA 93106, USA}}
\affiliation{\mbox{International Centre for Theoretical Sciences, Tata Institute of Fundamental Research, Bangalore 560089, India}}
 
% Because hyperref only gets the *last* author, we need to be explicit.
\hypersetup{pdfauthor={Islam et al.}}

\date{\today}

%==========================================================================
\begin{abstract}
Using publicly available numerical relativity (NR) simulations for non-spinning eccentric binary black hole (BBH) mergers, Ref \cite{Islam:2024rhm} demonstrated that the eccentricity-induced modulations in the amplitudes and frequencies of different spherical harmonic modes are mutually consistent and can be modeled using a single time series modulation. 
We extend the validity of the results to all non-precessing binaries by using 83 NR simulations from the SXS, RIT, and MAYA catalogs for aligned-spin eccentric BBH mergers with mass ratios ranging from $1:1$ to $1:4$. 
Based on these phenomenological relations, we provide a framework named \texttt{gwNRXHME} to compute multi-modal eccentric non-precessing waveforms using two inputs: quadrupolar eccentric waveforms, and the corresponding multi-modal quasi-circular non-precessing waveforms. 
Furthermore, we compute an overall degree of departure in SXS, RIT, and MAYA NR data from these relations and find that SXS NR simulations generally adhere to these relations more strictly than RIT and MAYA data. 
We also show that these relations can offer a cost-effective way to filter out noisy higher-order spherical harmonic modes extracted from NR data. 
Our framework is publicly available through the \texttt{gwModels} package.
\end{abstract}

\maketitle

%%%%%%%%%%%%%%%%%%%%%%%%%%%%%%%%%%%%%%%%%%%%%%%%%%%%%%%%%%%%%%%%%%%%%%%%%%%%%%%%%%%
%%%%%%%%%%%%%%%%%%%%%%%%%%%%%%%%%%%%%%%%%%%%%%%%%%%%%%%%%%%%%%%%%%%%%%%%%%%%%%%%%%%
\section{Introduction}
\label{sec:intro}
%%%%%%%%%%%%%%%%%%%%%%%%%%%%%%%%%%%%%%%%%%%%%%%%%%%%%%%%%%%%%%%%%%%%%%%%%%%%%%%%%%%
%%%%%%%%%%%%%%%%%%%%%%%%%%%%%%%%%%%%%%%%%%%%%%%%%%%%%%%%%%%%%%%%%%%%%%%%%%%%%%%%%%%
The detection and subsequent characterization of gravitational wave (GW) signals (radiation) from merging compact binaries in the LIGO-Virgo-KAGRA (LVK) data~\cite{Harry_2010,LIGOScientific:2014pky,VIRGO:2014yos,KAGRA:2020tym} depend on the availability of accurate numerical relativity (NR) waveforms and computationally efficient waveform models that are tuned to (or built from) NR data. 
Different waveform models adopt various simplifying assumptions about the underlying binary system in order to reduce the problem's dimensionality and facilitate the modeling process.

The gravitational waveform from a binary black hole (BBH) merger is expressed as a superposition of $-2$ spin-weighted complex-valued spherical harmonic modes with indices $(\ell, m)$~\cite{Maggiore:2007ulw,Maggiore:2018sht}:
\begin{align}
h(t, \theta, \phi; \boldsymbol{\lambda}) &= \sum_{\ell=2}^\infty \sum_{m=-\ell}^{\ell} h_{\ell m}(t; \boldsymbol\lambda)  _{-2}Y_{\ell m}(\theta,\phi),
\label{hmodes}
\end{align}
where $h$ is the complexified waveform constructed from the individual polarizations, $t$ represents time, $\theta$ and $\phi$ are angles on the merger's sky, and \bl is the set of intrinsic parameters that describe the binary.
If we use geometric units in which time is measured in units of the total mass of the binary, for non-precessing binaries, we have $\boldsymbol{\lambda}:=\{q,\chi_1,\chi_2,e_{\rm ref},l_{\rm ref}\}$, where $q:=m_1/m_2$ is the mass ratio, with $m_1$ and $m_2$ being the masses of the larger and smaller black holes, respectively. The parameter $\chi_1$ ($\chi_2$) denotes the dimensionless spin magnitudes of the larger (smaller) black hole. Eccentricity is characterized by two parameters: $e_{\rm ref}$, the eccentricity, and $l_{\rm ref}$, the mean anomaly, estimated at a chosen reference time or frequency. Multiple definitions of eccentricity and mean anomaly exist~\cite{Mroue:2010re, Healy:2017zqj,Shaikh:2023ypz,Knee:2022hth,2023PhRvD.108l4063R}; however, any consistent choice applied to all binaries will effectively characterize the eccentricity. 
We decompose each spherical harmonic mode $h_{\ell m}(t; \boldsymbol\lambda)$ into a real-valued amplitude $A_{\ell m}(t)$ and phase $\phi_{\ell m}(t)$ such that:
\begin{equation}
h_{\ell m}(t;\boldsymbol{\lambda}) = A_{\ell m}(t) e^{i \phi_{\ell m}(t)}.
\label{eq:amp_phase}
\end{equation}
We obtain the instantaneous frequency of each spherical harmonic mode as:
\begin{equation}
\omega_{\ell m}(t;\boldsymbol{\lambda}) = \frac{d\phi_{\ell m}(t)}{dt}.
\label{eq:freq}
\end{equation}
The orbital angular frequency of the binary is $\omega_{\text{orb}}=0.5 \times \omega_{22}$, and the dimensionless frequency is given by $x=(M \omega_{\text{orb}})^{2/3}$. 
In this paper, we only focus on the positive $m$ modes as the negative $m$ modes are obtained by the symmetry relation: $h_{\ell m} = (-1)^{\ell} h^{*}_{\ell-m}$ where $*$ indicates complex conjugate (this breaks down in precessing binaries \cite{2024PhRvD.109f3012T}). 

\begin{figure}
\includegraphics[width=\columnwidth]{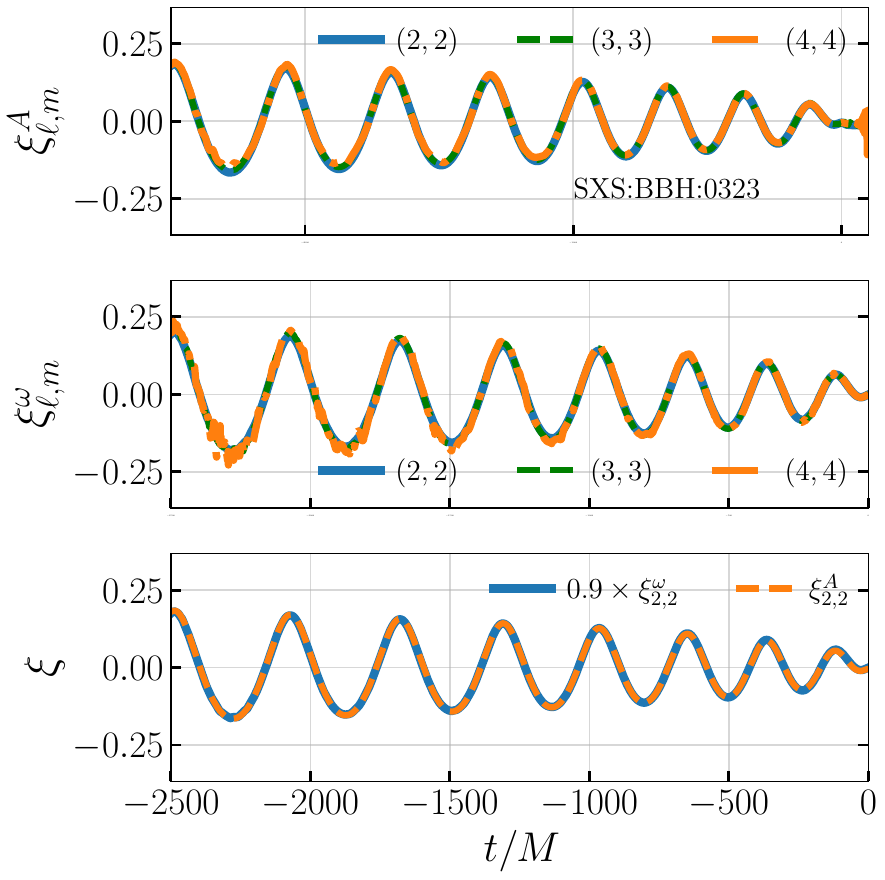}
\caption{We show the eccentric modulations in amplitudes $\xi_{\ell,m}^{A}$ (upper panel) and in frequencies $\xi_{\ell, m}^{\omega}$ (middle panel) for three representative modes: $(2,2)$ (blue) $(3,3)$ (green) and $(4,4)$ (orange) for a binary with mass ratio $q=1.22$, spin on the primary $\chi_1=0.33$, spin on the secondary $\chi_2=-0.44$ and eccentricity $e_{\rm ref}=0.193$. We extract this modulations from the eccentric NR simulation \texttt{SXS:BBH:0323} and the corresponding circular simulation \texttt{SXS:BBH:0318}. In the lower panel, we demonstrate that these two modulations are related by a factor of $K=0.9$ (obtained through a phenomenological fit provided in Ref.~\cite{Islam:2024rhm}).}
\label{fig:SXS0323}
\end{figure}

Years of effort have resulted in accurate waveform models for quasi-circular binaries~\cite{husa2016frequency,khan2016frequency,london2018first,khan2019phenomenological,
hannam2014simple,khan2020including,Pratten:2020ceb,Garcia-Quiros:2020qpx,Estelles:2020osj,Estelles:2020twz,Estelles:2021gvs,Estelles:2021gvs,Hamilton:2021pkf,bohe2017improved,cotesta2018enriching,cotesta2020frequency,pan2014inspiral, Pan:2009wj,Ghosh:2023mhc,Hamilton:2023znn,babak2017validating,cotesta2020frequency,pan2014inspiral,babak2017validating,Ossokine:2020kjp,Damour:2014sva,Nagar:2019wds,Nagar:2020pcj,Riemenschneider:2021ppj,Khalil:2023kep,Pompili:2023tna,Ramos-Buades:2023ehm,vandeMeent:2023ols,Blackman:2015pia,Blackman:2017dfb,Blackman:2017pcm,varma2019surrogate,varma2019surrogate2,Islam:2021mha}, but modeling radiation from eccentric binaries is still in its nascent stage~\cite{Tiwari:2019jtz, Huerta:2014eca, Moore:2016qxz, Damour:2004bz, Konigsdorffer:2006zt, Memmesheimer:2004cv,Hinder:2017sxy, Cho:2021oai,Chattaraj:2022tay,Hinderer:2017jcs,Cao:2017ndf,Chiaramello:2020ehz,Albanesi:2023bgi,Albanesi:2022xge,Riemenschneider:2021ppj,Chiaramello:2020ehz,Ramos-Buades:2021adz,Liu:2023ldr,Huerta:2016rwp,Huerta:2017kez,Joshi:2022ocr,Setyawati:2021gom,Wang:2023ueg,Islam:2021mha,Carullo:2023kvj,Nagar:2021gss,Tanay:2016zog}. Often, simple phenomenological relations observed in gravitational waveforms (either in the amplitudes or phases of different modes~\cite{london2018first}, or between precessing and non-precessing waveforms~\cite{Hamilton:2021pkf,khan2020including}) greatly aid the development of fast yet accurate waveform models. Recently, using publicly available NR simulations obtained from the SXS and RIT catalog, Ref.~\cite{Islam:2024rhm} provided a simple phenomenological relation between the amplitudes and instantaneous frequencies of different spherical harmonic modes in eccentric non-spinning binaries (i.e. with $\chi_1=\chi_2=0$), thereby effectively simplifying the process of waveform modeling for these binaries. 
\begin{figure}
\includegraphics[width=\columnwidth]{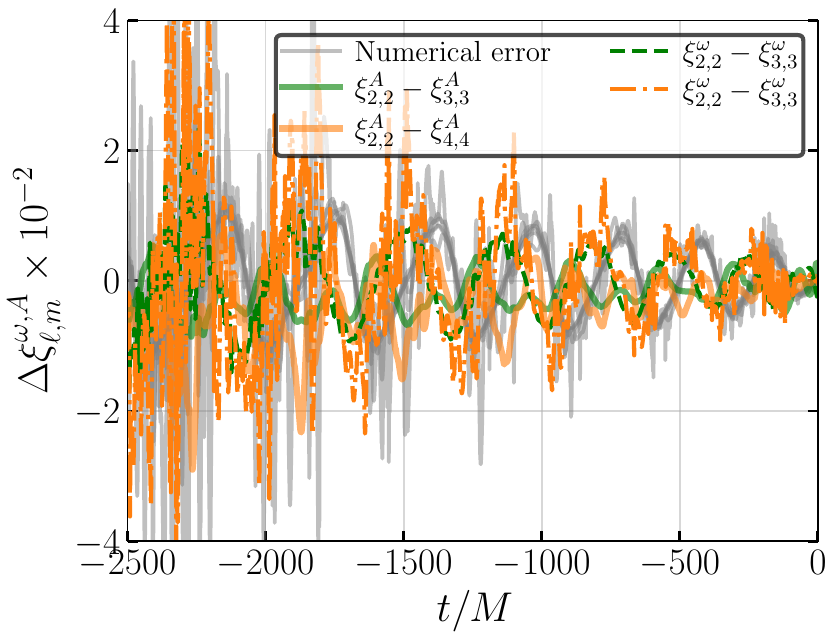}
\caption{We show the differences between eccentric amplitude modulations $\xi_{\ell,m}^{A}$ (solid lines; cf. Eq.(\ref{eq:amp_mod})) and frequency modulations $\xi_{\ell, m}^{\omega}$ (dashed lines; cf. Eq.(\ref{eq:freq_mod})) obtained from different modes (and shown in Figure~\ref{fig:SXS0323}) for a binary with mass ratio $q=1.22$, spin on the primary $\chi_1=0.33$, spin on the secondary $\chi_2=-0.44$ and eccentricity $e_{\rm ref}=0.193$. We extract these modulations from the eccentric NR simulation \texttt{SXS:BBH:0323} and the corresponding circular simulation \texttt{SXS:BBH:0318}. Additionally, we provide an estimate of numerical errors in both eccentric amplitude modulations and frequency modulations using the highest two resolutions of NR data (grey lines). We find that the differences between the modulations obtained from different modes are comparable to the numerical error in NR.}
\label{fig:SXS0323_modulations_error}
\end{figure}
Ref.~\cite{Islam:2024rhm} first quantifies the effect of eccentricity on different spherical harmonic modes in the eccentric waveform $h_{\ell m}(t;\boldsymbol{\lambda})$ by computing modulations with respect to the corresponding non-eccentric waveform $h_{\ell m}(t;\boldsymbol{\lambda}^0)$. Here, $\boldsymbol{\lambda}^0:=\{q,\chi_1,\chi_2,e_{\rm ref}=0,l_{\rm ref}=0\}$. Ref.~\cite{Islam:2024rhm} defines the eccentric frequency modulation as:
\begin{equation}
\xi_{\ell m}^{\omega}(t;\boldsymbol{\lambda}) = b_{\ell m}^\omega \frac{\omega_{\ell m}(t;\boldsymbol{\lambda})-\omega_{\ell m}(t;\boldsymbol{\lambda^0})}{\omega_{\ell m}(t;\boldsymbol{\lambda^0})},
\label{eq:freq_mod}
\end{equation}
and the eccentric amplitude modulation as:
\begin{equation}
\xi_{\ell m}^{A}(t;\boldsymbol{\lambda}) = b^{A}_{\ell m} \frac{2}{\ell} 
\frac{A_{\ell m}(t;\boldsymbol{\lambda})-A_{\ell m}(t;\boldsymbol{\lambda^0})}{A_{\ell m} (t;\boldsymbol{\lambda^0)}},
\label{eq:amp_mod}
\end{equation}
where $b_{\ell m}^\omega\! =b^{A}_{\ell m}\!=1$. Ref.~\cite{Islam:2024rhm} then empirically noted that (i) the amplitude modulations $\xi_{\ell m}^{A}(t;\boldsymbol{\lambda})$ are consistent for all the modes used; (ii) the frequency modulations $\xi_{\ell m}^{\omega}(t;\boldsymbol{\lambda})$ obtained from different modes are all consistent; and
(iii) the amplitude modulations and frequency modulations are related by: 
\begin{equation}
\label{eq:amp_freq_mod_relation}
\xi_{\ell m}^{A}(t;\boldsymbol{\lambda}) = B \, \xi_{\ell m}^{\omega}(t;\boldsymbol{\lambda}),
\end{equation}
where the scaling factor $B = 0.9$. Note that while $\xi_{\ell m}^{\omega}(t;\boldsymbol{\lambda})$ does not have any mode dependent pre-factor, the pre-factor in $\xi_{\ell m}^{A}(t;\boldsymbol{\lambda})$ depends on the $\ell$ value of the spherical harmonic mode.
In this paper, we use 83 publicly available NR simulations for non-precesing eccentric BBH mergers (see Table~\ref{tab:sxsdata_entries}, Table~\ref{tab:ritdata_entries} and Table~\ref{tab:mayadata_entries}) from three different NR catalogs, SXS catalog (\href{https://data.black-holes.org/waveforms/catalog.html}{https://data.black-holes.org/waveforms/catalog.html})~\cite{Boyle:2019kee,Hinder:2017sxy,LIGOScientific:2016ebw}, RIT catalog (\href{https://ccrgpages.rit.edu/~RITCatalog/}{https://ccrgpages.rit.edu/~RITCatalog/}~\cite{Healy:2022wdn,Healy:2020vre}, and MAYA catalog (\href{https://cgp.ph.utexas.edu/waveform}{https://cgp.ph.utexas.edu/waveform}~\cite{Ferguson:2023vta}), which employ different codes to solve the Einstein equations, to probe the validity of the aforementioned relations in aligned-spin binaries. 

\begin{table}[h!]
\centering
\footnotesize % or \footnotesize for even smaller text
\setlength{\tabcolsep}{10pt} % Increase space between text and column boundary
\begin{tabular}{l c r}
 &  &  \\
\toprule
SXS:BBH:0108 & SXS:BBH:1355 & SXS:BBH:1366 \\
SXS:BBH:0319 & SXS:BBH:1356 & SXS:BBH:1367 \\
SXS:BBH:0320 & SXS:BBH:1357 & SXS:BBH:1368 \\
SXS:BBH:0321 & SXS:BBH:1358 & SXS:BBH:1371 \\
SXS:BBH:0322 & SXS:BBH:1359 & SXS:BBH:1372 \\
SXS:BBH:0323 & SXS:BBH:1364 & SXS:BBH:1373 \\
SXS:BBH:1149 & SXS:BBH:1365 & SXS:BBH:1169 \\
\toprule
\end{tabular}
\caption{SXS NR simulations~\cite{Hinder:2017sxy} used in this work.}
\label{tab:sxsdata_entries}
\end{table}

\begin{table}[h!]
\centering
\footnotesize % or \footnotesize for even smaller text
\setlength{\tabcolsep}{10pt} % Increase space between text and column boundary
\begin{tabular}{l c r}
 &  &  \\
\toprule
RIT:eBBH:1899 & RIT:eBBH:1099 & RIT:eBBH:1254 \\
RIT:eBBH:1900 & RIT:eBBH:1100 & RIT:eBBH:1256 \\
RIT:eBBH:1828 & RIT:eBBH:1282 & RIT:eBBH:1330 \\
RIT:eBBH:1807 & RIT:eBBH:1210 & RIT:eBBH:1353 \\
RIT:eBBH:1786 & RIT:eBBH:1213 & RIT:eBBH:1376 \\
RIT:eBBH:1763 & RIT:eBBH:1215 & RIT:eBBH:1399 \\
RIT:eBBH:1740 & RIT:eBBH:1422 & RIT:eBBH:1445 \\
RIT:eBBH:1883 & RIT:eBBH:1252 & RIT:eBBH:1468 \\
RIT:eBBH:1862 & RIT:eBBH:1253 & RIT:eBBH:1491 \\
RIT:eBBH:1098 & & \\
\toprule
\end{tabular}
\caption{RIT NR simulations~\cite{Healy:2022wdn,Healy:2020vre} used in this work.}
\label{tab:ritdata_entries}
\end{table}

\begin{table}[h!]
\centering
\footnotesize % or \footnotesize for even smaller text
\setlength{\tabcolsep}{10pt} % Increase space between text and column boundary
\begin{tabular}{l c r}
 &  &  \\
\toprule
MAYA0952 & MAYA0916 & MAYA0965 \\
MAYA0953 & MAYA0917 & MAYA0966 \\
MAYA0974 & MAYA0918 & MAYA0967 \\
MAYA0975 & MAYA0919 & MAYA0968 \\
MAYA0976 & MAYA0941 & MAYA0969 \\
MAYA0977 & MAYA0942 & MAYA0985 \\
MAYA0995 & MAYA0943 & MAYA0986 \\
MAYA0996 & MAYA0944 & MAYA0987 \\
MAYA0997 & MAYA0945 & MAYA0988 \\
MAYA0998 & MAYA0963 & MAYA0989 \\
MAYA0999 & MAYA0964 & MAYA0990 \\
MAYA0991 & & \\
\toprule
\end{tabular}
\caption{MAYA NR simulations~\cite{Ferguson:2023vta} used in this work.}
\label{tab:mayadata_entries}
\end{table}

\begin{figure*}[htb]
\centering
\subfigure[]{\label{fig:RIT1862}
\includegraphics[scale=0.56]{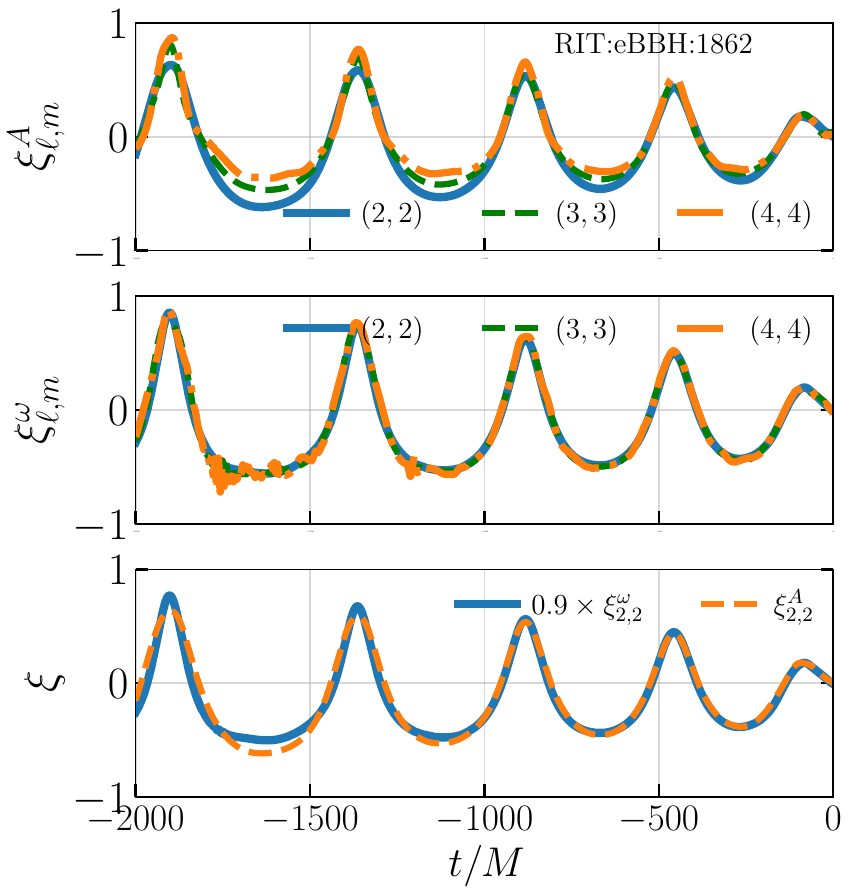}}
\subfigure[]{\label{fig:MAYA0969}
\includegraphics[scale=0.56]{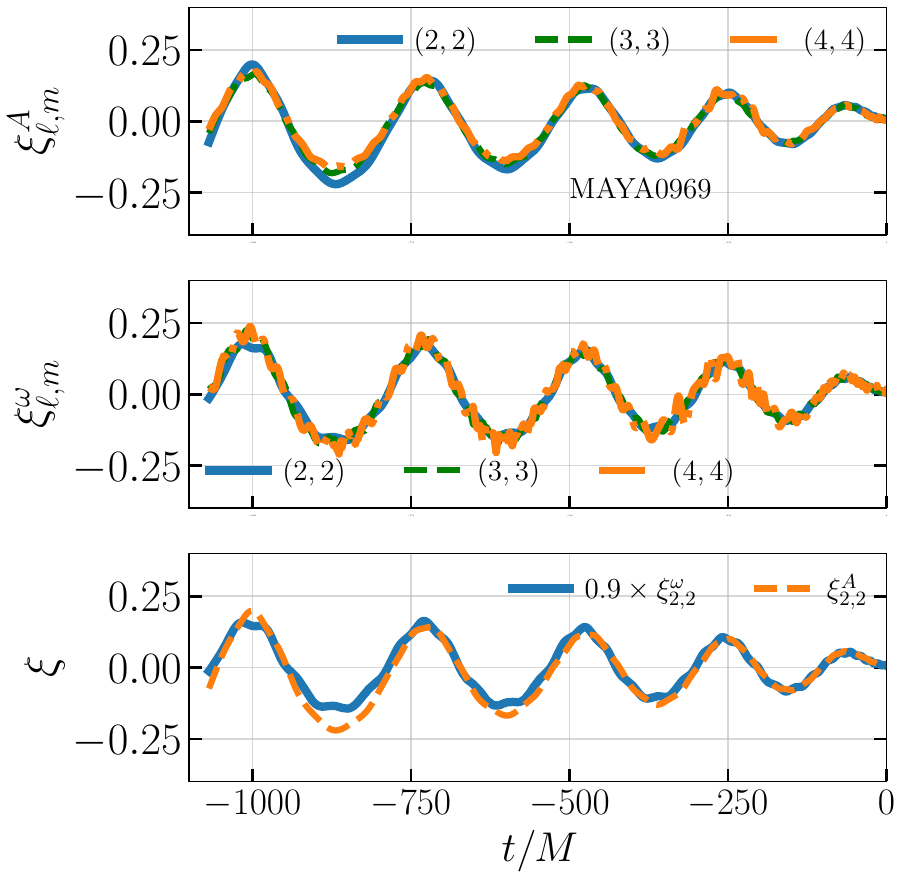}} 
\caption{(a) We show the eccentric modulations in amplitudes $\xi_{\ell,m}^{A}$ (upper panel) and in frequencies $\xi_{\ell, m}^{\omega}$ (middle panel) for three representative modes: $(2,2)$ (blue) $(3,3)$ (green) and $(4,4)$ (orange) for a binary with mass ratio $q=3$, spin on the primary $\chi_1=-0.8$, spin on the secondary $\chi_2=0.0$ and eccentricity $e_{\rm ref}=0.36$~\cite{Healy:2022wdn,Healy:2020vre}. We extract this modulations from the eccentric NR simulation \texttt{RIT:eBBH:1862} and the corresponding circular waveform is obtained from \texttt{NRHybSur3dq8} model. In the lower panel, we demonstrate that these two modulations are related by a factor of $K=0.9$ (obtained through a phenomenological fit provided in Ref.~\cite{Islam:2024rhm}). (b) We show the same quantities for  the eccentric NR simulation \texttt{MAYA0969} characterized by mass ratio $q=3$, spin on the primary $\chi_1=0.4$, spin on the secondary $\chi_2=0.4$ and eccentricity $e_{\rm ref}=0.112$~\cite{Ferguson:2023vta}.}
\end{figure*}

%%%%%%%%%%%%%%%%%%%%%%%%%%%%%%%%%%%%%%%%%%%%%%%%%%%%%%%%%%%%%%%%%%%%%%%%%%%%%%%%%%%
%%%%%%%%%%%%%%%%%%%%%%%%%%%%%%%%%%%%%%%%%%%%%%%%%%%%%%%%%%%%%%%%%%%%%%%%%%%%%%%%%%%
\section{Quasi-universal phenomenological relation}
%%%%%%%%%%%%%%%%%%%%%%%%%%%%%%%%%%%%%%%%%%%%%%%%%%%%%%%%%%%%%%%%%%%%%%%%%%%%%%%%%%%
%%%%%%%%%%%%%%%%%%%%%%%%%%%%%%%%%%%%%%%%%%%%%%%%%%%%%%%%%%%%%%%%%%%%%%%%%%%%%%%%%%%
We first demonstrate the relations mentioned in Section~\ref{sec:intro} for an aligned-spin eccentric BBH merger waveform obtained from NR simulation \texttt{SXS:BBH:0323} (Figure~\ref{fig:SXS0323}). This binary is characterized by mass ratio $q=1.22$, spin on the primary $\chi_1=0.33$, spin on the secondary $\chi_2=-0.44$ and eccentricity $e_{\rm ref}=0.193$ (as obtained from the simulation metadata provided at \href{https://data.black-holes.org/waveforms/catalog.html}{https://data.black-holes.org/waveforms/catalog.html}). The corresponding quasi-circular BBH merger simulation is \texttt{SXS:BBH:0318}. First, we show that the amplitude modulations $\xi_{\ell m}^{A}$ obtained from different spherical harmonic modes are consistent (Figure~\ref{fig:SXS0323}, upper panel). Second, the frequency modulations $\xi_{\ell m}^{\omega}$ in different spherical harmonic modes also match with each other (Figure~\ref{fig:SXS0323}, middle panel). Finally, amplitude and frequency modulations are related by the factor $B$ (Figure~\ref{fig:SXS0323}, lower panel). 
For each spherical harmonic mode, we compute the differences between $\xi_{\ell m}^{A}(t;\boldsymbol{\lambda})$ and $\xi_{\ell m}^{\omega}(t;\boldsymbol{\lambda})$ computed from NR simulations with different numerical resolution. This gives an estimate of the numerical errors in the simulations. For the highest resolution NR simulation, we then compute the differences in $\xi_{\ell m}^{A}(t;\boldsymbol{\lambda})$ obtained from the $(2,2)$, $(3,3)$ and $(4,4)$ modes. We do the same for $\xi_{\ell m}^{\omega}(t;\boldsymbol{\lambda})$. These differences quantify the degree of dispersion about the phenomenological relations. Figure~\ref{fig:SXS0323_modulations_error} shows that where the differences between $\xi_{\ell m}^{A}(t;q,e_{\rm ref})$ (and $\xi_{\ell m}^{\omega}(t;q,e_{\rm ref})$) computed from different spherical harmonic modes closely follow the numerical errors in NR simulations.

Next, we show the quasi-universal eccentric modulations obtained from one of the representative RIT NR BBH merger simulations: \texttt{RIT:eBBH:1862}, characterized by $q=3$, $\chi_1=-0.8$, $\chi_2=0.0$, and $e_{\rm ref}=0.36$~\cite{Healy:2022wdn,Healy:2020vre}, in Figure~\ref{fig:RIT1862}. The corresponding quasi-circular waveforms are obtained from \texttt{NRHybSur3dq8}~\cite{varma2019surrogate}, an accurate reduced-order surrogate approximation of the aligned-spin SXS NR data. This choice was made because, for the non-precessing eccentric NR simulations in the RIT and MAYA catalogs, we could not always find a corresponding quasi-circular simulation. Figure~\ref{fig:MAYA0969} shows the same quantities observed in \texttt{MAYA0969}, a representative MAYA NR BBH merger simulations. This particular binary has the following parameters: $q=3$, $\chi_1=0.4$, $\chi_2=0.4$, and $e_{\rm ref}=0.112$~\cite{Ferguson:2023vta}. It is important to stress that while it is possible to characterize these eccentric simulations using a common set of definitions in this paper, it is not necessary to understand the universal relations observed. We therefore provide the eccentricity values as noted in the respective NR catalog papers.
The relations between the various $\xi_{\ell m}^{A}$ (and $\xi_{\ell m}^{\omega}$), as well as the relation between $\xi_{\ell m}^{A}$ and $\xi_{\ell m}^{\omega}$ are qualitatively apparent in the RIT and MAYA simulations as well (Figures~\ref{fig:RIT1862},~\ref{fig:MAYA0969}), but overall the relations are not obeyed at the level that they are in the SXS simulation (Figure~\ref{fig:SXS0323}). 
Note that the modulations in Figures~\ref{fig:RIT1862} and \ref{fig:MAYA0969} show a higher level of numerical noise, which is more prominent for the $(3,3)$ and $(4,4)$ modes. 
The level of numerical noise in many of the other RIT and MAYA NR simulations considered in this paper is larger than what is shown in these figures.

\begin{figure}
\includegraphics[width=\columnwidth]{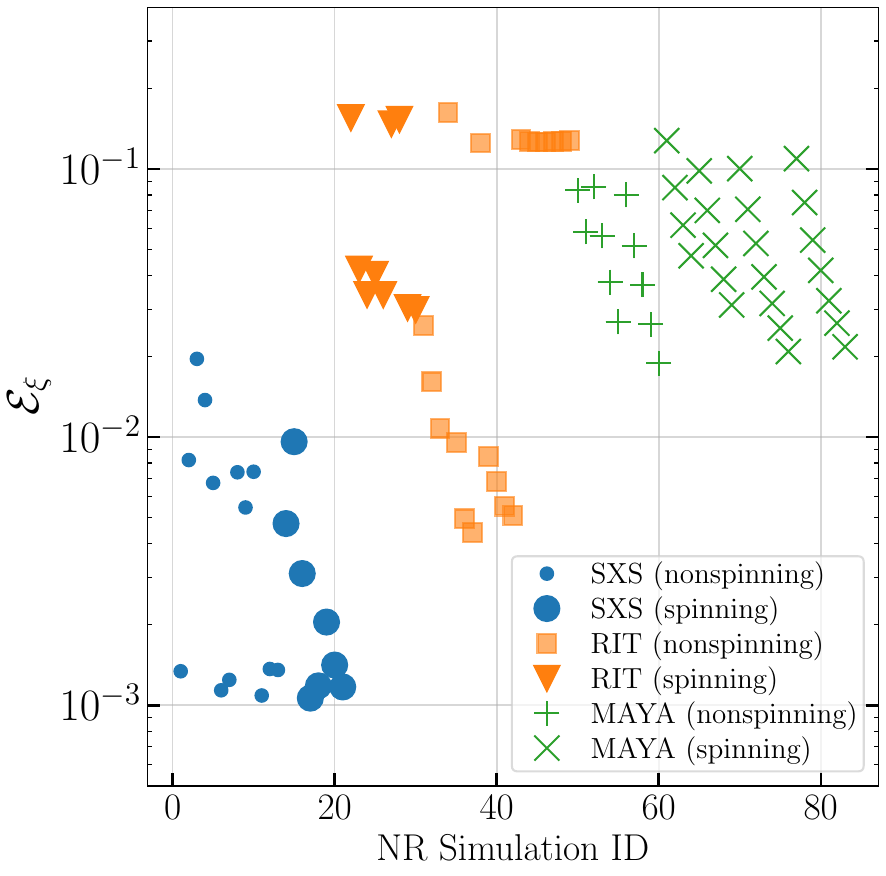}
\caption{We show the overall degree of departures $\mathcal{E}_{\xi}$ (defined in Eq.(\ref{eq:overall_err})) associated with the quasi-universal relations (presented in Section~\ref{sec:intro}) observed in all non-precessing NR simulations considered in this work. Different markers are used for non-spinning and spinning waveforms from the SXS, RIT, and MAYA catalogs.}
\label{fig:Modelling_errors}
\end{figure}

\begin{figure}
\includegraphics[width=\columnwidth]{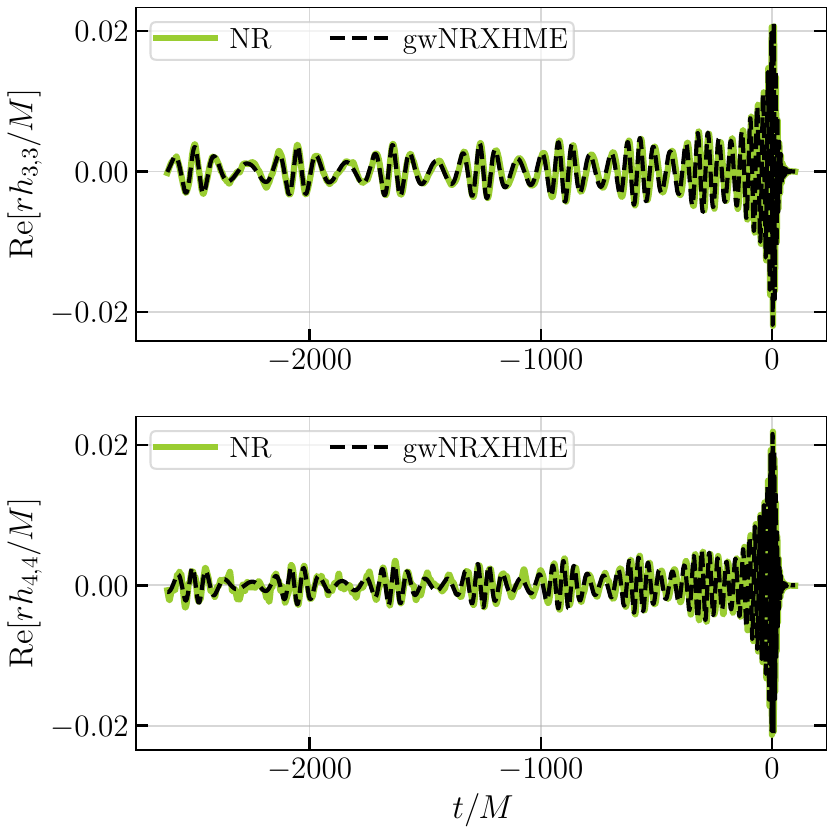}
\caption{We show the eccentric higher-order spherical harmonic modes (black dashed lines) obtained from the \texttt{gwNRXHME} framework (available at \href{https://github.com/tousifislam/gwModels}{https://github.com/tousifislam/gwModels}) and corresponding NR data from \texttt{SXS:BBH:0324} simulation (green solid lines). We obtain \texttt{gwNRHME} predictions by applying eccentric modulations (computed from the $(2,2)$ mode of \texttt{SXS:BBH:0324} data) on quasi-circular spherical harmonic modes obtained from \texttt{SXS:BBH:0318}. We find that \texttt{gwNRXHME} predictions are visually indistinguishable from NR.}
\label{fig:SXS0324_NRHMEcc_wfs}
\end{figure}

Finally, we compute the degree of departure from the presented phenomenological relations, from Eq.(\ref{eq:freq_mod}) to Eq.(\ref{eq:amp_freq_mod_relation}), for all NR simulations considered in this work. We use the relative $L_2$-norm to quantify the degree of departure. For two given time-series $s_1(t)$ and $s_2(t)$, the relative $L_2$-norm is defined as (Eq.21 of Ref.~\cite{Blackman:2017dfb}):
\begin{equation}
\mathcal{E}(s_1,s_2) = \frac{1}{2} \frac{\int_{t_{\rm min}}^{t_{\rm max}}|s_{1}(t) - s_{2}(t)|^2 dt}{\int_{t_{\rm min}}^{t_{\rm max}}|s_{1}(t)|^2 dt},
\label{eq:l2err}
\end{equation}
where $t_{\rm min}$ and $t_{\rm max}$ denote the initial and final times, respectively. 

To ensure that our estimates for the degree of departure account for (i) differences in $\xi_{\ell m}^{A}$ estimated from different modes, (ii) differences in $\xi_{\ell m}^{\omega}$ estimated from different modes, and (iii) differences in the proposed relation between $\xi_{\ell m}^{A}$ and $\xi_{\ell m}^{\omega}$, we define an overall degree of departure $\mathcal{E}_{\xi}$ associated with the quasi-universal relations, presented in Section~\ref{sec:intro}, as follows:
\begin{subequations}
\begin{align}
\mathcal{E}_1 &= \mathcal{E}(\xi_{22}^{A},\xi_{33}^{A}),
\\
\mathcal{E}_2 &= \mathcal{E}(\xi_{22}^{A},\xi_{44}^{A})
\\
\mathcal{E}_3 &= \mathcal{E}(\xi_{22}^{\omega},\xi_{33}^{\omega}).
\\
\mathcal{E}_4 &= \mathcal{E}(\xi_{22}^{\omega},\xi_{44}^{\omega}).
\\
\mathcal{E}_5 &= \mathcal{E}(\xi_{22}^{A},B\xi_{22}^{\omega}).
\\
\mathcal{E}_{\xi} &= \frac{\mathcal{E}_1+\mathcal{E}_2+\mathcal{E}_3+\mathcal{E}_4+\mathcal{E}_5}{5}.  
\end{align}
\label{eq:overall_err}
\end{subequations}
These five errors are not really independent of each other, so we define the net error as an average of these five errors to give a sense of the typical error due to one of those terms.
For equal mass binaries, odd $m$ modes are zero due to symmetry in the system. We therefore do not compute $\mathcal{E}_3$ and $\mathcal{E}_5$ in those cases. The overall degree of departure is also modified accordingly as: $\mathcal{E}_{\xi}=\frac{\mathcal{E}_1+\mathcal{E}_2+\mathcal{E}_4}{3}$.

We show the overall degree of departure $\mathcal{E}_{\xi}$ computed for all non-precessing NR simulations in Figure~\ref{fig:Modelling_errors}. Different markers are used for non-spinning and spinning waveforms from the SXS, RIT, and MAYA catalogs. The spinning RIT data exhibit larger overall degree of departures compared to the non-spinning cases. For non-spinning cases, the overall degree of departures in RIT data are somewhat comparable to those in SXS data. However, SXS NR simulations generally adhere to these relations more strictly than RIT and MAYA data. For the SXS data, the overall degree of departures associated with these universal relations mostly lies between $10^{-3}$ and $10^{-2}$, whereas overall errors for the MAYA simulations range between $10^{-2}$ and $10^{-1}$. For the RIT simulations, errors range from $0.004$ to $0.2$. Motivated by the level of numerical scatter in the empirically calculated modulations in Figures~\ref{fig:RIT1862} and~\ref{fig:MAYA0969}, we conjecture that the larger departures from the phenomenological relations in the RIT and MAYA simulations may be related to the numerical errors in the simulations, but this fact requires further investigation. Note that some of the SXS NR simulations too exhibit increased level of numerical noises - mostly in the higher order modes. These simulations corresponds to the largest overall errors. These errors provide quantitative and qualitative evidences in support of the existence of a quasi-universal relation between spherical harmonic modes in non-precessing eccentric BBH waveforms. Furthermore, it provides a new benchmark to compare eccentric NR simulations obtained from different catalog and codes.

%%%%%%%%%%%%%%%%%%%%%%%%%%%%%%%%%%%%%%%%%%%%%%%%%%%%%%%%%%%%%%%%%%%%%%%%%%%%%%%%%%%
%%%%%%%%%%%%%%%%%%%%%%%%%%%%%%%%%%%%%%%%%%%%%%%%%%%%%%%%%%%%%%%%%%%%%%%%%%%%%%%%%%%
\section{Implications}
%%%%%%%%%%%%%%%%%%%%%%%%%%%%%%%%%%%%%%%%%%%%%%%%%%%%%%%%%%%%%%%%%%%%%%%%%%%%%%%%%%%
%%%%%%%%%%%%%%%%%%%%%%%%%%%%%%%%%%%%%%%%%%%%%%%%%%%%%%%%%%%%%%%%%%%%%%%%%%%%%%%%%%%
Our results have several interesting implications.
First, it extends the validity of the quasi-universal relations observed earlier between modes in non-spinning eccentric BBH merger waveforms~\cite{Islam:2024rhm} to cases involving aligned-spin and anti-aligned-spin configurations. Our results show that these relations are observed, with varying accuracy, in NR data obtained from three different catalogs.
Second, as a consequence, the results provide a novel method for comparing current and future eccentric NR catalogs by assessing how well they adhere to these quasi-universal relations. 

Third, as noted in Ref.~\cite{Islam:2024rhm}, these relations greatly simplify the modeling challenges in eccentric binaries. They indicate that one can combine dominant quadrupolar mode of a non-precessing eccentric (NR) waveform $h_{\ell m}(t;\boldsymbol{\lambda})$ with the corresponding quasi-circular non-precessing multi-modal (NR) waveform $h_{\ell m}(t;\boldsymbol{\lambda}^0)$ to obtain multi-modal non-precessing eccentric waveforms. It offers an alternate and simpler approach to filter out the otherwise noisy higher-order modes in eccentric NR simulations. For this, one has to simply exploit Eq.(\ref{eq:freq_mod}), Eq.(\ref{eq:amp_mod}) and Eq.(\ref{eq:amp_freq_mod_relation}). Ref.~\cite{Islam:2024rhm} has presented such a framework named  \texttt{gwNRHME} for the non-spinning eccentric binaries. Our result shows that the framework can be easily extended for the non-precessing binaries. We call our new framework \texttt{gwNRXHME}. 

\begin{figure}
\includegraphics[width=\columnwidth]{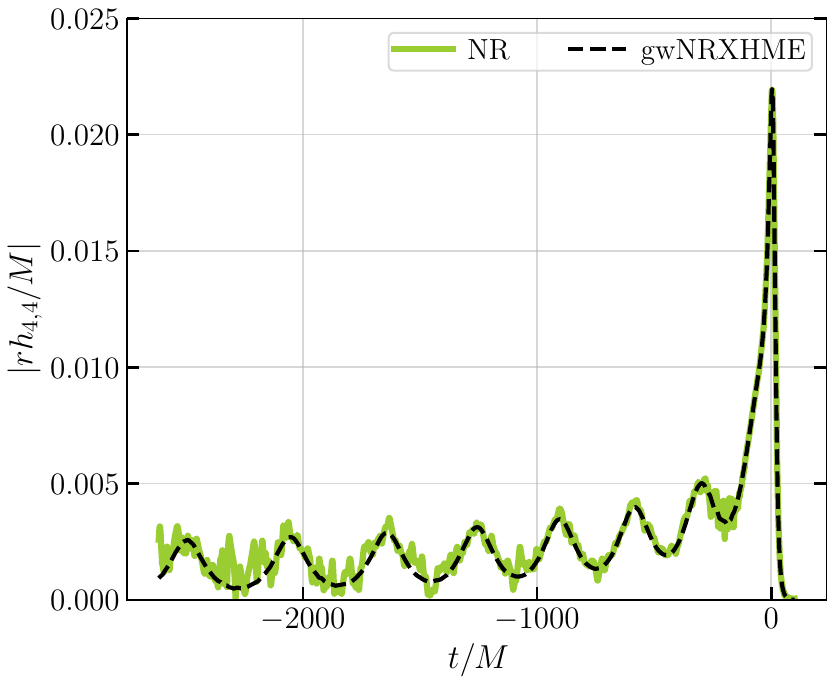}
\caption{We demonstrate that \texttt{gwNRXHME} framework (black dashed line; available at \href{https://github.com/tousifislam/gwModels}{https://github.com/tousifislam/gwModels}) can be used to provide clean eccentric higher-order spherical harmonic modes. We illustrate this for \texttt{SXS:BBH:0324} (green), characterized by a mass ratio $q=1.22$, spins $\chi_1=0.33$ and $\chi_2=-0.4$ and eccentricity $e_{\rm ref}=0.13$~\cite{LIGOScientific:2016ebw}. While the amplitude of the $(\ell,m)=(4,4)$ mode obtained from NR is quite noisy, the \texttt{gwNRXHME} prediction (obtained by combining $(2,2)$ mode eccentric modulation from \texttt{SXS:BBH:0324} simulation with quasi-circular \texttt{SXS:BBH:0318} simulation) is clean and goes right through the NR data.}
\label{fig:smooth_NRHME}
\end{figure}

We first define an common modulation parameter $\xi(t) := \xi_{22}^{A}(t; \boldsymbol{\lambda})$.
The amplitude of higher order modes is given as:
\begin{equation}
A_{\ell m}^{\tt{gwNRXHME}}(t; \boldsymbol{\lambda}) = A_{\ell m}(t; \boldsymbol{\lambda}^0) \Big[ \frac{\ell}{2} \xi(t) + 1 \Big].
\end{equation}
The frequency of different spherical harmonic modes is then simply:
\begin{equation}
\omega_{\ell m}^{\tt{gwNRXHME}}(t; \boldsymbol{\lambda}) = \omega_{\ell m}(t; \boldsymbol{\lambda}^0) \Big[ \frac{\xi(t)}{B} + 1 \Big].
\end{equation}
Integrating the frequency, we get the phase of each mode:
\begin{equation}
\phi_{\ell m}^{\tt{gwNRXHME}}(t; \boldsymbol{\lambda}) = \phi_{0} + \int \omega_{\ell m}^{\tt{gwNRHME}}(t; \boldsymbol{\lambda}) dt,
\end{equation}
where $\phi_{0}=\phi_{\ell m}(t; \boldsymbol{\lambda}^0)$ is the integration constant. Finally, We obtain complex time-series for each mode as:
\begin{equation}
h_{\ell m}^{\tt{gwNRXHME}}(t; \boldsymbol{\lambda}) = A_{\ell m}^{\tt{gwNRXÍHME}}(t; \boldsymbol{\lambda}) e^{\phi_{\ell m}^{\tt{gwNRXHME}}(t;\boldsymbol{\lambda})}.
\end{equation}
We demonstrate this framework using non-precessing eccentric NR simulation \texttt{SXS:BBH:0324} and corresponding quasi-circular NR simulation \texttt{SXS:BBH:0318}. Both simulations are for $q=1.22$, $\chi_1=0.33$ and $\chi_2=-0.44$. Simulation \texttt{SXS:BBH:0324} is characterized by eccentricity of $e_{\rm ref}=0.13$ measured at a reference frequency of $\omega=0.017351$~\cite{LIGOScientific:2016ebw}.
In Figure~\ref{fig:SXS0324_NRHMEcc_wfs}, we show the eccentric higher-order spherical harmonic modes (black dashed lines) obtained from the \texttt{gwNRXHME} framework and corresponding NR data from \texttt{SXS:BBH:0324} simulation (green solid lines). We find that \texttt{gwNRXHME} predictions are visually indistinguishable from NR. In Figure~\ref{fig:smooth_NRHME}, we focus on the $(4,4)$ mode amplitude. We illustrate that 
while the NR data is noisy, \texttt{gwNRXHME} prediction for the $(4,4)$ mode exhibits noise-free amplitude and accurately tracks the noisy NR data. 

As an immediate next step, we can use the \texttt{gwNRXHME} framework to combine existing multi-modal quasi-circular non-precessing waveform models, such as \texttt{NRHybSur3dq8}, \texttt{IMRPhenomTHM}, or \texttt{SEOBNRv5HM}, with an existing quadrupolar mode eccentric waveform model to develop accurate multi-modal non-precessing eccentric waveform models. This approach will significantly reduce the complexity and time required to build multi-modal non-precessing eccentric waveform models. We would like to mention that such exercise has already been performed for non-spinning eccentric binaries in Ref.~\cite{Islam:2024zqo} using \texttt{gwNRHME} framework. We should also note that these phenomenological relations have so far been tested only for non-precessing BBH merger scenarios. It is necessary to study their validity in precessing eccentric scenarios. However, few NR simulations for precessing eccentric BBH mergers exist, making it difficult to perform such tests. Therefore, dedicated NR simulations for precessing eccentric BBH mergers are needed, which we leave for future work.

%%%%%%%%%%%%%%%%%%%%%%%%%%%%%%%%%%%%%%%%%%%%%%%%%%%%%%%%%%%%%%%%%%%%%%%%%%%%%%%%%%%%%%%%%%%%%%%%%%%%%%%%
%%%%%%%%%%%%%%%%%%%%%%%%%%%%%%%%%%%%%%%%%%%%%%%%%%%%%%%%%%%%%%%%%%%%%%%%%%%%%%%%%%%%%%%%%%%%%%%%%%%%%%%%
\begin{acknowledgments}
We are grateful to the SXS collaboration, RIT NR group and MAYA collaboration for maintaining publicly available catalog of NR simulations which has been used in this study. We thank Scott Field, Gaurav Khanna, Deborah Ferguson, Vijay Varma, Chandra Kant Mishra, Prayush Kumar and Saul Teukolsky for helpful discussions, and Shrobana Ghosh for comments and suggestions on the initial draft.
This research was supported in part by the National Science Foundation under Grant No. NSF PHY-2309135 and the Simons Foundation (216179, LB). 
TV acknowledges support from NSF grants 2012086 and 2309360, the Alfred P. Sloan Foundation through grant number FG-2023-20470, the BSF through award number 2022136, and the Hellman Family Faculty Fellowship. Use was made of computational facilities purchased with funds from the National Science Foundation (CNS-1725797) and administered by the Center for Scientific Computing (CSC). The CSC is supported by the California NanoSystems Institute and the Materials Research Science and Engineering Center (MRSEC; NSF DMR 2308708) at UC Santa Barbara.
\end{acknowledgments}
%%%%%%%%%%%%%%%%%%%%%%%%%%%%%%%%%%%%%%%%%%%%%%%%%%%%%%%%%%%%%%%%%%%%%%%%%%%%%%%%%%%%%%%%%%%%%%%%%%%%%%%%
%%%%%%%%%%%%%%%%%%%%%%%%%%%%%%%%%%%%%%%%%%%%%%%%%%%%%%%%%%%%%%%%%%%%%%%%%%%%%%%%%%%%%%%%%%%%%%%%%%%%%%%%

%%%%%%%%%%%%%%%%%%%%%%%%%%%%%%%%%%%%%%%%%%%%%%%%%%%%%%%%%%%%%%%%%%%%%%%%%%%%%%%%%%%%%%%%%%%%%%%%%%%%%%%%
\bibliography{References}
%%%%%%%%%%%%%%%%%%%%%%%%%%%%%%%%%%%%%%%%%%%%%%%%%%%%%%%%%%%%%%%%%%%%%%%%%%%%%%%%%%%%%%%%%%%%%%%%%%%%%%%%

\end{document}